\documentclass[allclo]{FBSart} 
\usepackage{amsfonts}
\usepackage{amssymb}
\usepackage{epsfig}

\title{Energy Dependence of the Delta Resonance: Chiral Dynamics in
Action} 
\author{A. M. Bernstein\instnr{1} and S. Stave\instnr{2}}

\instlist{Physics Department and Laboratory for Nuclear Science, 
MIT, Cambridge, MA
02139, USA 
\and
Department of Physics, Duke University, Durham,
  North Carolina 27708, USA and Triangle Universities Nuclear
  Laboratory, Durham, North Carolina 27708, USA}  

\runningauthor{A.\,M.\,Bernstein} 
\runningtitle{Energy Dependence of the Delta Resonance: Chiral Dynamics in
Action} 

\begin{document}

\maketitle

\begin{abstract}
There is an important connection between the low energy theorems of QCD and the energy dependence of the  $\Delta$ resonance in $\pi$-N scattering, as well as the closely related  $\gamma^{*} N \rightarrow N \pi$ reaction. The resonance shape is due not only to the strong $\pi$-N interaction in the p wave but the small interaction in the s wave; the latter  is due to spontaneous chiral symmetry breaking in QCD (i.e. the Nambu-Goldstone nature of the pion). A brief overview of experimental tests of  chiral perturbation theory and chiral based models  is presented.\\
\end{abstract}

\section{Introduction}\label{sec:intro}
  Since the discovery of the $\Delta$ resonance\footnote{The modern
    values for the $\Delta$ are $I(J^{p}$)=3/2(3/2$^{+}$), center of
    mass energy $W$ = 1232 MeV, width $\Gamma$ =118 MeV\cite{PDB}.} (the first excited state of the nucleon) by Fermi's group\cite{Fermi} it has been well known  that it dominates  low energy $\pi N$ scattering and the closely related $\gamma^{*} N \rightarrow \pi N$ reaction. In the intervening years there has been a great deal of experimental and theoretical activity  in this classical field of $\pi N$ physics. 
This is  central to our understanding of nuclei (the nucleon-nucleon
potential) and of the long range properties of hadrons through their
virtual emission and absorption of pions. In this article we shall
stress the less well known relationship between the spontaneous hiding
of chiral symmetry in QCD, its subsequent low energy theorems,  and
the energy dependence of the $\Delta$ resonance. A brief overview of low energy $\pi$N physics is presented with an emphasis on electromagnetic pion production.  It is shown that theoretical calculations based on spontaneous chiral symmetry hiding  economically summarize the wealth of accurate
data that have been taken in the past two decades\cite{CD2006,BM-reviews,AB-CD2006}. We present this as
a tribute to S.~N. Yang who has been a leader in using a chiral based pion cloud model (the Dubna-Mainz-Taipei or DMT model\cite{DMT}) to successfully  predict the observables in these low energy reactions. 
  
  The QCD Lagrangian can be written as a sum of two terms, $L_{0}$
  which is independent of the lightest quark masses (up, down) and
  $L_{m}$ which contains the masses of the  two light quarks\cite{DGH}. Consider
  the chiral limit in which the   light quark masses $m_{q}
  \rightarrow 0$.  As is well known, the vector current is conserved
  while the axial vector current is conserved only in the chiral limit
  (i.e. $m_{q} \rightarrow 0$) and slightly non-conserved in the real
  world. This is one of the approximate symmetries of QCD on which chiral perturbation theory (ChPT) is based\cite{CD2006,BM-reviews,DGH}.  Despite the fact that the light quark mass
  independent part of the QCD Lagrangian,  $L_{0}$, has chiral
  symmetry, matter does not seem to obey the rules. The chiral
  symmetry is expected to show up by the parity doubling of all
  hadronic states: i.e., the proton with $j^{p} =1/2^{+}$ would have a
  $1/2^{-}$ partner (the Wigner-Weyl manifestation of the symmetry). Clearly, this is not the case. This indicates that the symmetry is spontaneously hidden (often  stated as spontaneously broken) and is manifested in the Nambu-Goldstone  mode; the parity doubling occurs through  the appearance of a massless pseudo scalar ($0^{-}$) meson. The opposite parity partner of the proton is a proton and a ``massless pion'' (Goldstone Boson). The consequence of this for the  $\pi N$ interaction in momentum space is:
  
  \begin{equation}\label{eq:VpiN}
   V_{\pi N} = g_{\pi N} \vec{\sigma} \cdot \vec{p_{\pi}}
   \end{equation} 
   where $\vec{\sigma}$ is the nucleon spin and $\vec{p_{\pi}}$ is the pion momentum. In
accordance with Goldstone's theorem, this interaction  $\rightarrow 0$
as the pion momentum $\rightarrow 0$. Furthermore, the coupling constant  $g_{\pi N}$ can be computed from the Goldberger-Treiman relation\cite{DGH} and chiral corrections\cite{Goity:GT} and is accurate to the few \% level. The $\pi N$  interaction is very weak in the s wave and strong in the p
wave which leads to the $\Delta$ resonance, the tensor force between
nucleons, and to long range non-spherical virtual pionic contributions
to hadronic structure. These salient features of the $\pi N$ interaction have been known for decades and can be found in most textbooks on nuclear physics. However, they are usually based on empirical findings such as the pseudoscalar nature of the pion and on the empirically determined coupling constant $g_{\pi N}$. What is different in this presentation is the fact that it is based on QCD and that these empirical findings are in fact predicted by the considerations of spontaneous chiral symmetry hiding in QCD. 

Equation~(\ref{eq:VpiN}) shows that the cross sections for $\pi N$
scattering must go to zero at low energies in the chiral limit. This
was first derived before the advent of QCD using current algebra (now
recognized as   the lowest order 
chiral perturbation theory (ChPT) calculation $O(p^{2}$)) for $a(\pi,h)$, the s wave $\pi$
hadron scattering length\cite{W:pion-scat}. The result is    $a^{I}(\pi,h) = - \vec I_{\pi} \cdot
\vec{I_{h} }m_{\pi}/(\Lambda_{x}  F_{\pi})$ where  $\vec{I} =
\vec{I_{\pi}} + \vec{I_{h}} $ is the total isospin, $\vec{I_{\pi}}$ and
$\vec{I_{h}}$ are the isospin of  the pion and hadron respectively,
$F_{\pi}$ is the pion decay constant, and $\Lambda_{x} = 4 \pi F_{\pi}
\simeq $ 1 GeV is the chiral symmetry breaking
scale\cite{W:pion-scat}. Note that $a(\pi,h)\rightarrow  0$ in the
chiral limit, $ m_{\pi} \rightarrow  0$, as it must  to obey Goldstone's
theorem. Also note that  $a(\pi,h) \simeq 1/\Lambda_{x}  \simeq $ 0.1
fm, which is small compared to a typical strong interaction scattering
length of $\simeq$ 1 fm. This small scattering length  is obtained
from the explicit chiral symmetry breaking due to the finite quark
masses. The predictions of ChPT for $\pi$N scattering lengths have
been verified in detail in a beautiful series of experiments on pionic
hydrogen and deuterium at PSI\cite{Gotta:atoms}. 

Low energy electromagnetic production of Goldstone Bosons is as
fundamental as Goldstone Boson scattering for two   reasons: 1) the
production amplitudes vanish in the chiral limit (as  in scattering);
and 2) the phase of the production amplitude is linked to  scattering
in the final state by unitarity or the final state interaction
(Fermi-Watson) 
theorem suitably modified to take the up, down quark masses
into account\cite{AB:FW}.  First consider the low energy limit of the
electric dipole $E_{0+}$ for s wave photo-pion
production\cite{BKM-photo}:
\begin{equation}
\begin{array}{rcl}
E_{0+}(\gamma p \rightarrow \pi^{0}p) & = &- D_{0} \mu ( 1 +
O(\mu)+..)\rightarrow 0\\
E_{0+}(\gamma p \rightarrow \pi^{+}n) & = & \sqrt{2} D_{0}/(1+ \mu 
+...)^{3/2} \rightarrow \sqrt{2} D_{0}\\
\mu & = & m_{\pi}/M \rightarrow 0\\
D_{0} =e \cdot g_{\pi N}/8 \pi M& = & 24 \cdot 10^{-
3 }(1/m_{\pi})\\
\end{array}
\label{eq:E0}
\end{equation} 
where $M$ is the nucleon mass and the right arrow denotes
the chiral limit  ($m_{u}, m_{d},m_{\pi} \rightarrow
0$). Equation~(\ref{eq:E0}) shows that for neutral pion production the
amplitude vanishes in the chiral limit.  For charged pion  production,
there is a different low energy theorem\cite{BKM-photo}. Therefore, the
amplitude that is most sensitive to explicit chiral symmetry breaking
is  neutral pion production and most of the modern  experiments have
concentrated on this channel. In general, ChPT to one  loop calculated
in the heavy Fermion approximation has been highly  successful in
calculating the observed cross sections and linearly polarized photon
asymmetry\cite{BKM-photo}.

\section{$\pi N$ and $\gamma N \rightarrow \pi N$ Experiments}
\label{piN} 

\subsection{Energy Dependence of the $\Delta$ Resonance}
 
\begin{figure}
\begin{center} 
\epsfig{file=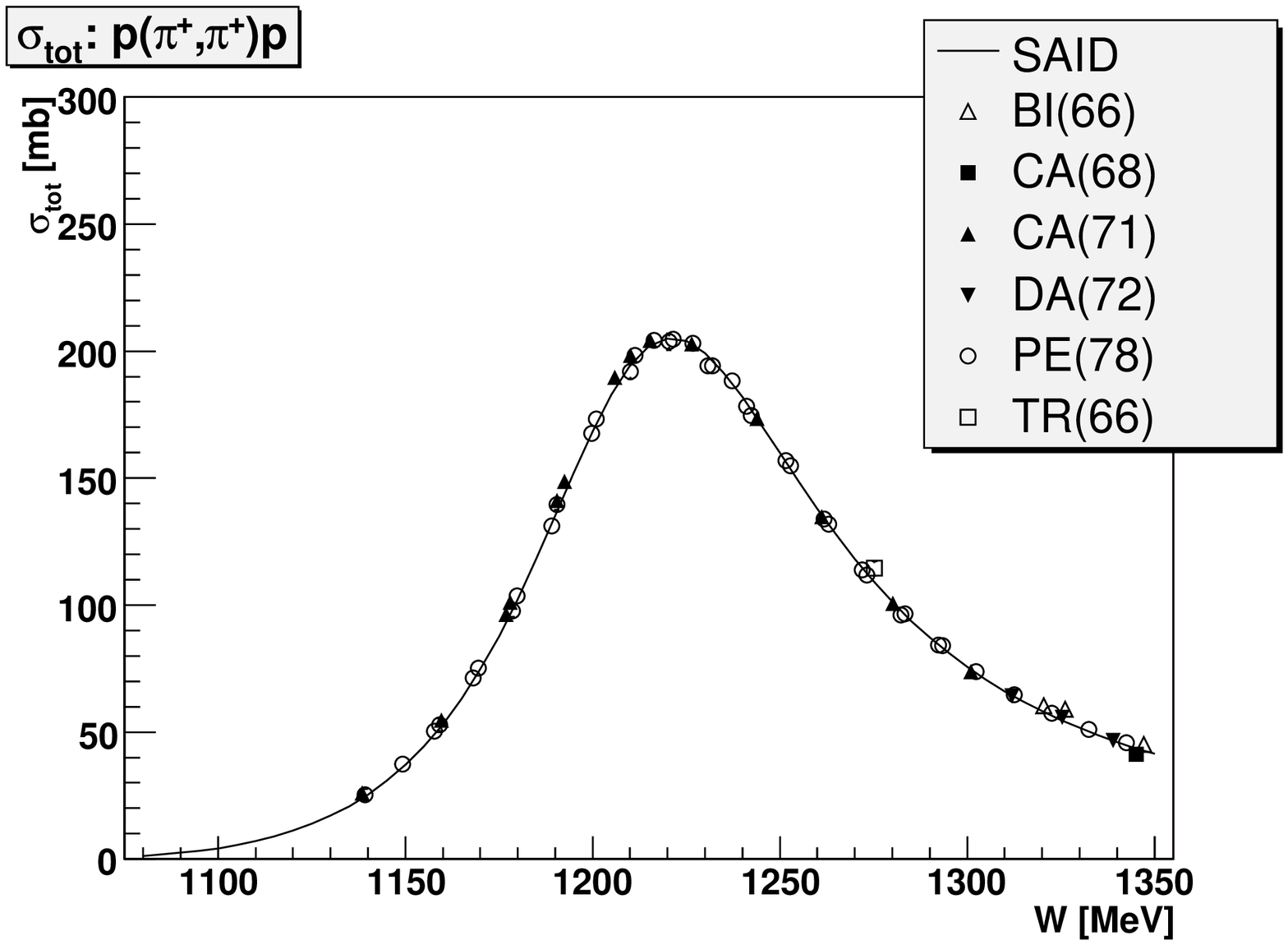, width=2.6in}
\epsfig{file=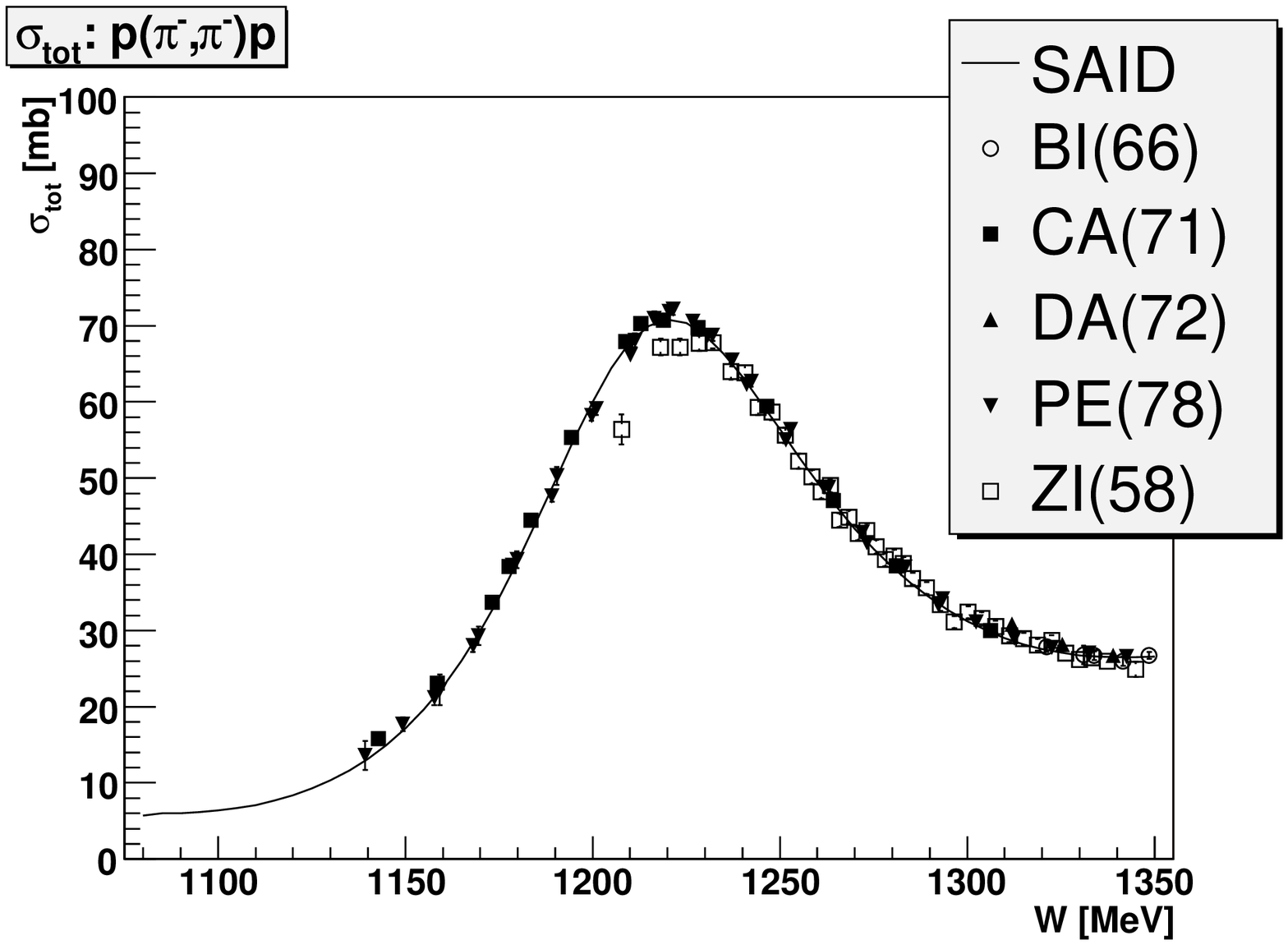,width=2.6in}
\end{center}
\caption{\label{fig:piN}  The total cross section for $\pi$N scattering at low energies   as a function of $W$, the center of mass energy, through the $\Delta$ resonance. 
 Left panel:  $\pi^{+}$p
 scattering. Right panel: $\pi^{-}$p scattering. The data and the fits are from the SAID compilation\cite{SAID}.   }
\end{figure}

\begin{figure}
\begin{center} \epsfig{file=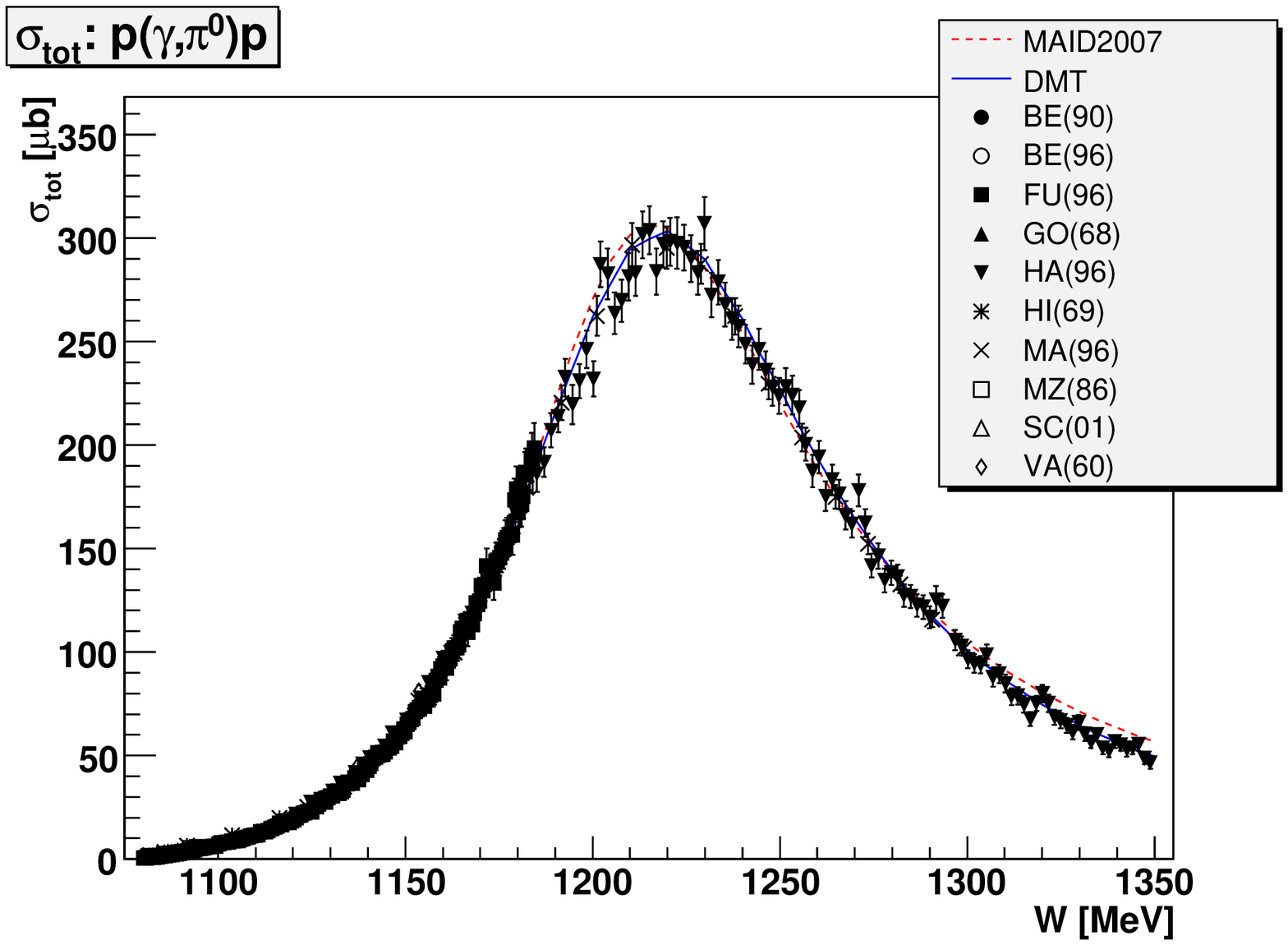, width=2.6in}
\epsfig{file=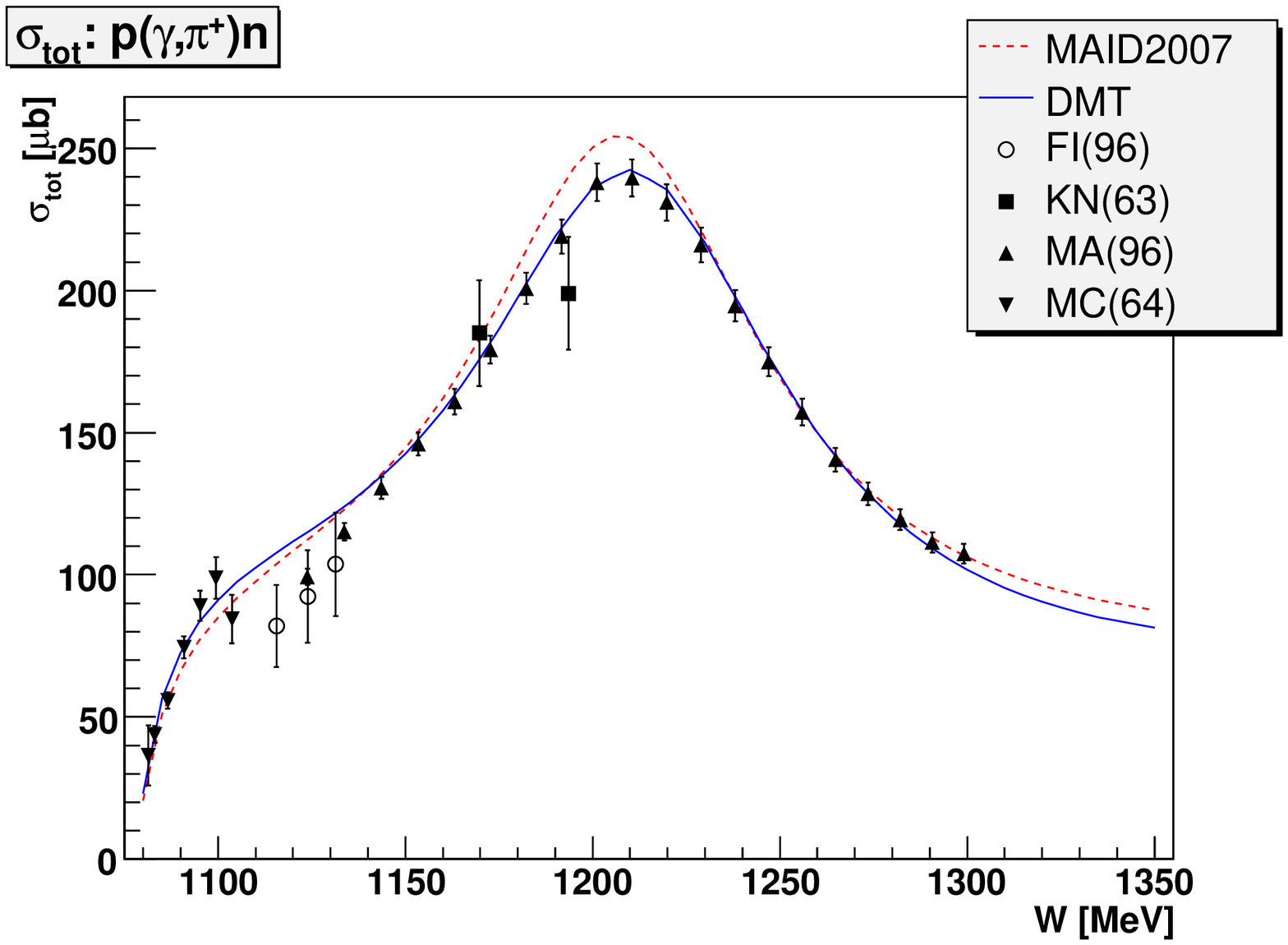,width=2.6in}
\end{center}
\caption{\label{fig:photo} The $\gamma p \rightarrow \pi N$ cross sections as a function of $W$, the center of mass energy, from threshold through the $\Delta$ region.  Left panel:  $\gamma p \rightarrow\pi^{0}p$. Right panel: $\gamma p \rightarrow \pi^{+} n$.  The data are from the SAID compilation\cite{SAID} and the curves represent the results of the DMT\cite{DMT} and MAID models\cite{MAID}.}  
\end{figure}

The application of the ideas of the previous section to data from low energy $\pi N$
scattering and electromagnetic pion production from the nucleon is
instructive. In this section we shall take a broad view of the energy
dependence of the $\pi N$ interaction from threshold through the
$\Delta$ resonance as revealed by total cross section data 
(amplifying a brief previous presentation\cite{AB-CD2006}). Figure~\ref{fig:piN} shows the  total cross sections for
$\pi^{+/-} p$  scattering\cite{SAID}. These reactions have a strong $\Delta$ resonance. As expected, the  $\pi^{+}p$ 
 cross section goes to zero near threshold. The small, but not zero, cross section for $\pi^{-}p$ scattering near threshold is due to Coulomb effects.  These two cross sections 
clearly show the $\Delta$ resonance without any interference (the
small shift between them is due to the mass difference of the
$\Delta^{0}$ and $\Delta^{+}$). Indeed these  cross sections are a
textbook 
example of an isolated resonance. Although not usually mentioned
in textbooks  it is the combination of a strong resonance and a small
cross section at threshold that produces this beautiful example (as
predicted by chiral dynamics)! This  can be verified experimentally in the case of photo-pion production shown in Fig.~\ref{fig:photo}. If we consider the $\gamma p \rightarrow \pi^{0} p$ reaction, the cross section near threshold goes to zero as indicated by Eq.~(\ref{eq:E0}) and the $\Delta$ resonance looks very similar to $\pi N$ scattering. On the other hand, for 
 the $\gamma p \rightarrow \pi^{+} n$ reaction there is  strong s wave production starting at threshold, due to the Kroll-Ruderman low energy theorem (see Eq.~(\ref{eq:E0})). In this case the $\Delta$ resonance curve is superimposed on the strong s wave amplitude and looks quite different! 
 
 From Figure~\ref{fig:photo} we see that the two model curves are in
 good agreement with the data. These are the phenomenological
 MAID\cite{MAID} and the  pion cloud DMT (Dubna-Mainz-Taipei)\cite{DMT} models, in which S.~N. Yang plays a major role. The reason for this good agreement with experiment is that both models have the low energy theorems of QCD as well as an accurate description of the $\Delta$ resonance. 

\begin{figure}
\begin{center} 
\epsfig{file=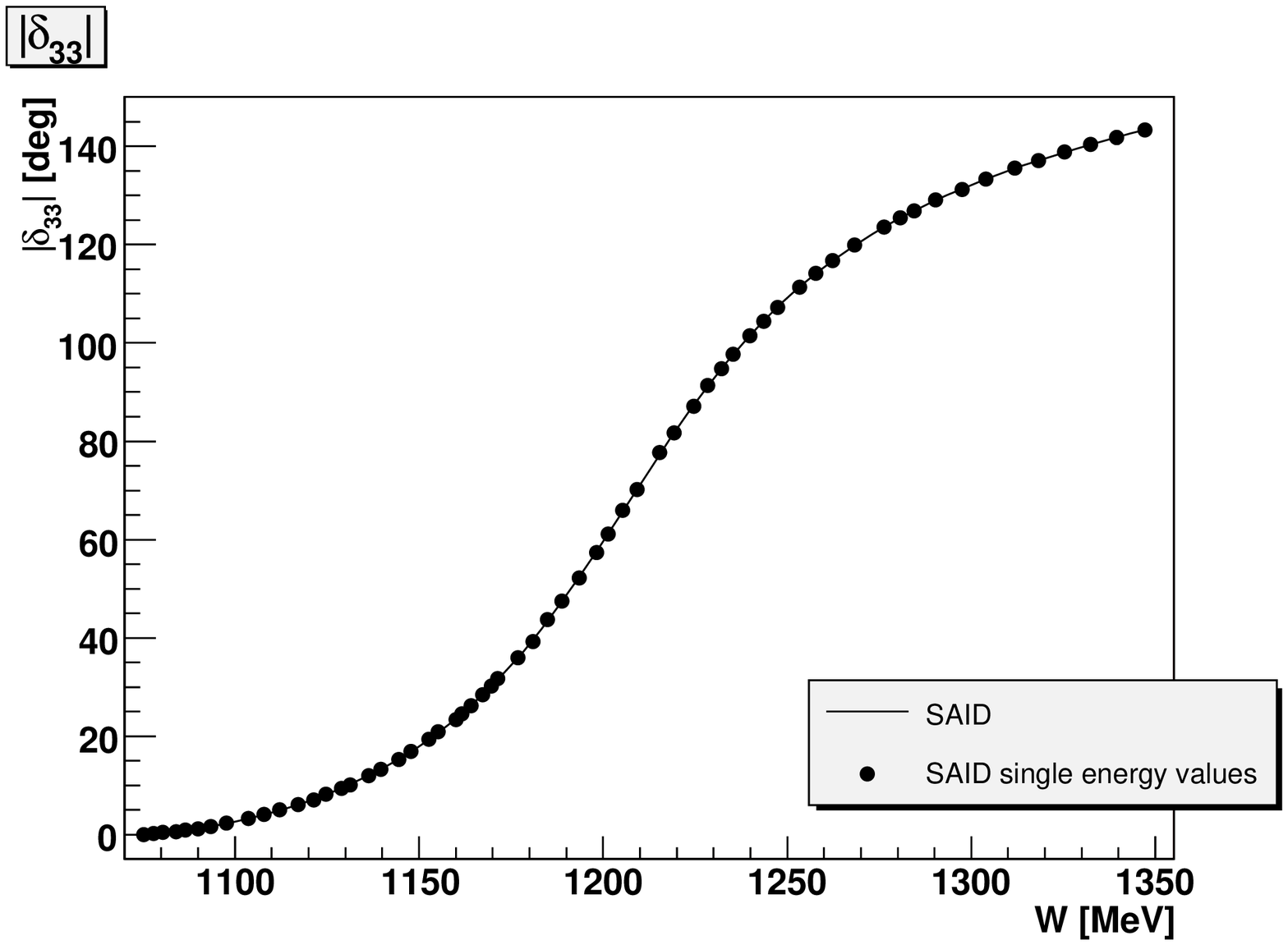,width=2.6in}
\epsfig{file=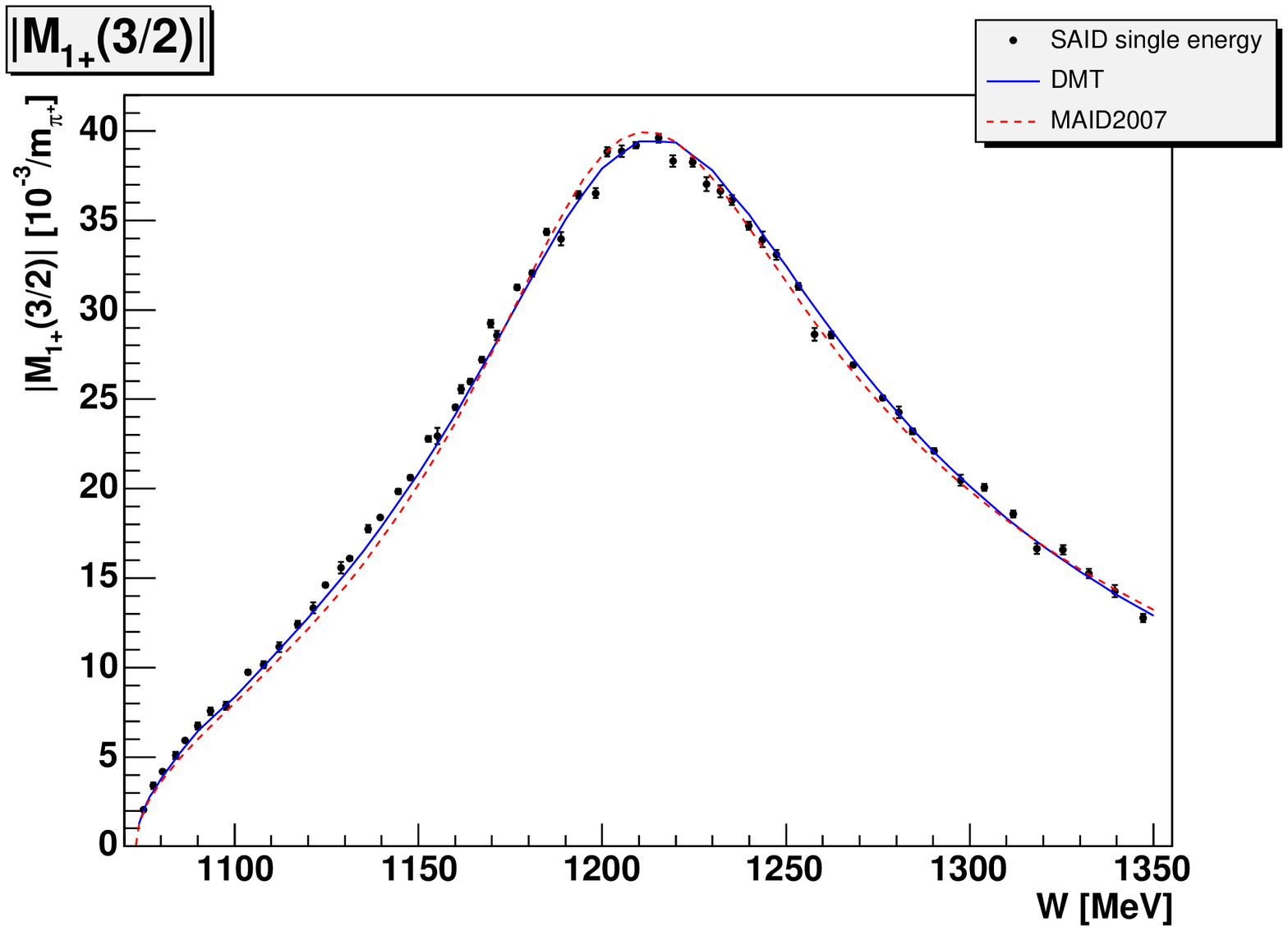, width=2.6in}
\end{center}

\caption{\label{fig:Delta}  Left panel: The $\delta_{33}$ phase shift for $\pi$N scattering versus $W$, the center of mass energy, from threshold through the $\Delta$ resonance. The points  are  the SAID single energy fits to the data and the curve is the smooth energy dependent fit\cite{SAID}. Right panel: The absolute value of the resonant  $M_{1+}(I=3/2)$ multipole 
for the $\gamma p \rightarrow \pi N$ reaction    as a function of $W$.
The points  are from the SAID single energy analysis\cite{SAID} and the curves are the results of the DMT\cite{DMT} and MAID\cite{MAID} models.  }
\end{figure}

The energy dependence of the $\Delta$ resonance can also be seen very clearly in the  $\delta_{33}$ ($I= J = 3/2$) phase shift in $\pi$N scattering and in the 
$M_{1+}(I=3/2)$ for the $\gamma^{*} N \rightarrow \pi N $ reaction
(for the notation see \cite{notation,obs}). These have the advantage
that they show the resonance directly. Since the observables are
bilinear combinations of the transition matrix elements neither the
phase shifts nor multipoles are directly observable. In general they
have been extracted from experiment by model dependent methods. In
this case, where we are exhibiting the dominant amplitudes,
 the model errors are believed to be small. In Figure~\ref{fig:Delta} we present these quantities from the SAID analysis\cite{SAID}. It can be seen that $\delta_{33}$  for $\pi$N scattering passes through 90$^{\circ}$ in the upwards direction at $W = 1232$ MeV which defines the $\Delta$ resonance position. The width comes from a Breit-Wigner fit to the energy dependence. The magnitude of the resonant $M_{1+}(I=3/2)$ amplitude for the $\gamma N \rightarrow \pi N $ reaction is also shown in Fig.~\ref{fig:Delta}. It also defines the same position and width for the $\Delta$ as does $\pi$N scattering. It can also be seen that the MAID\cite{MAID} and SAID\cite{SAID} models are in good agreement with this dominant photo-pion resonant amplitude.

\subsection{Tests of Theoretical Calculations}

In this section we give a brief overview of the present status of
experimental tests of chiral perturbation theory (ChPT) and pion cloud model calculations of the electromagnetic pion production for the threshold and $\Delta$ regions. 

\begin{figure}
\begin{center}
\epsfig{file=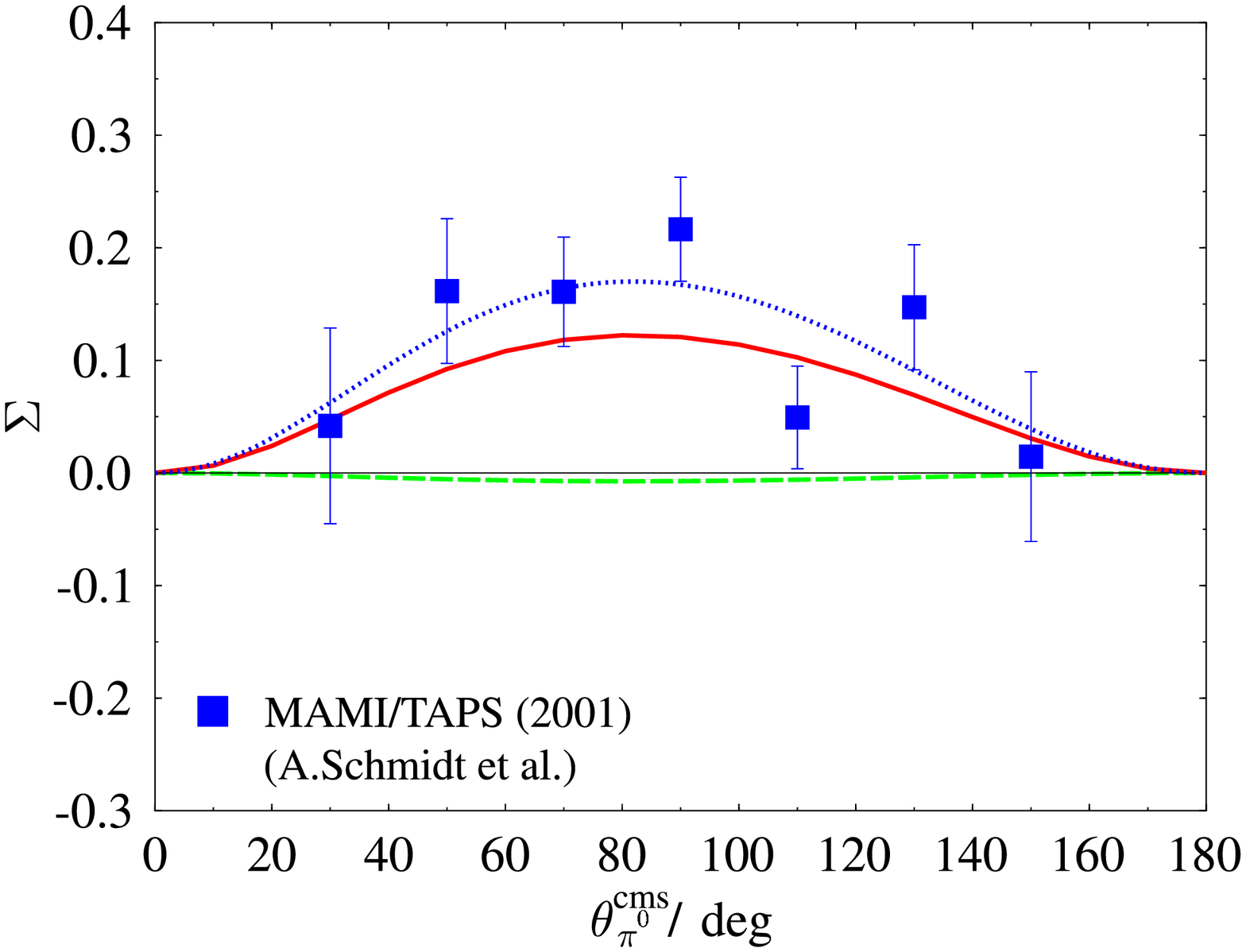,width=2.8in}
\epsfig{file=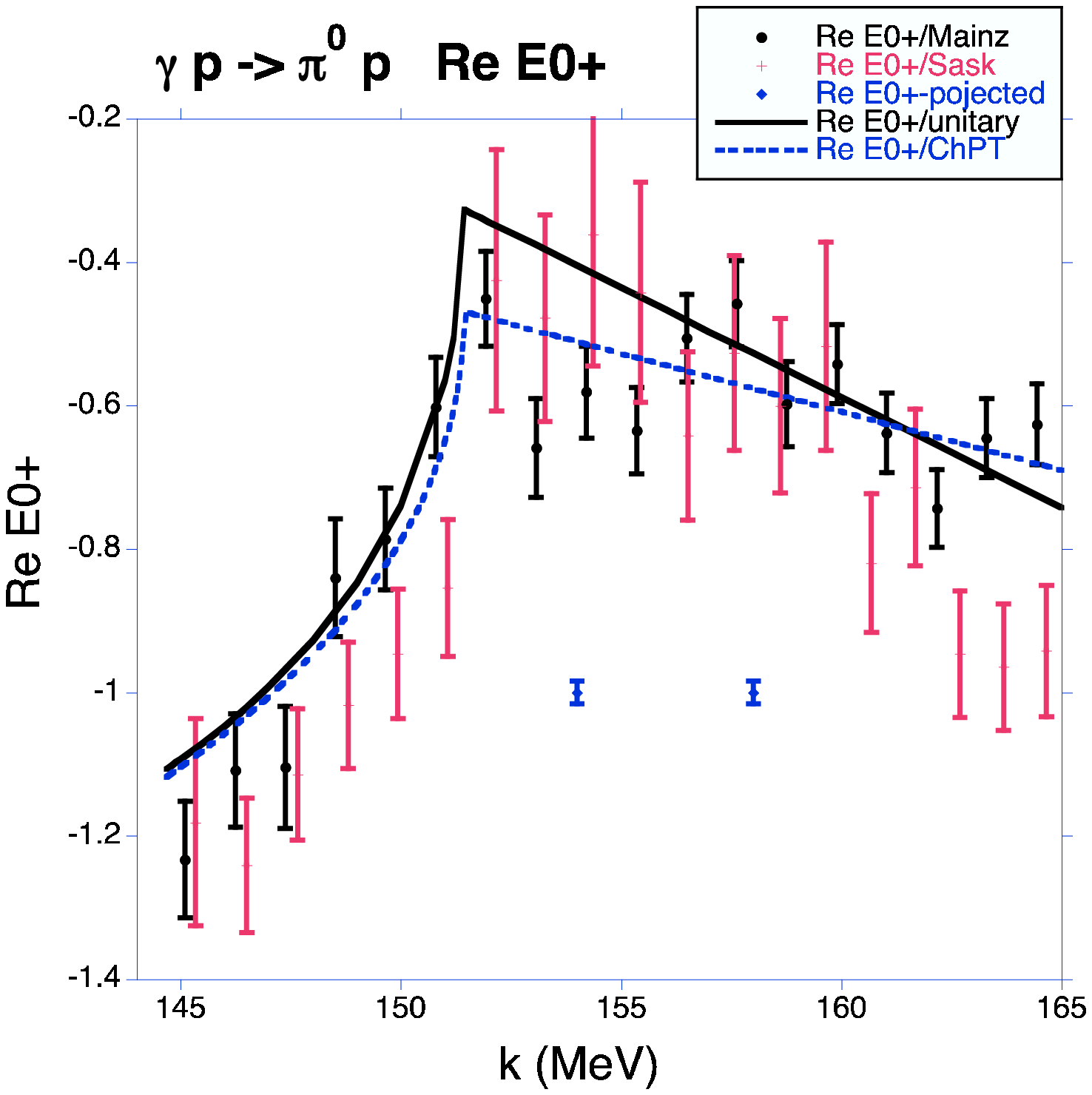, width=2.24in}
\end{center}
\caption{\label{fig:photo_thresh}
The $\gamma p \rightarrow \pi^{0} p$ reaction.  Left panel: Polarized photon asymmetry 
$\Sigma$ versus angle at an average photon energy of 159.5
MeV\cite{Schmidt}.  The solid (red) curve is ChPT, O($p^{3}$), the
dotted (blue) curve is ChPT, O($p^{4}$)\cite{BKM-photo}. The dashed
(green) curve is from dispersion theory\cite{dispersion-photo}.  
Right panel: $\Re E_{0+}$\cite{notation} versus photon energy. The data points are from Mainz\cite{Schmidt} and Saskatoon\cite{Sask}. The curves are from ChPT\cite{BKM-photo} and a unitary fit to the data\cite{AB:FW}. The two projected points from HI$\gamma$S  are plotted at an  arbitrary value ($\Re E_{0+}$ = -1)  to show the anticipated statistical errors for ~200 hours of running time per point. See text for discussion. }
\end{figure}

A great deal of effort has gone into the study of the near threshold
$\gamma p \rightarrow \pi^{0} p$ reaction experimentally at
Mainz\cite{Schmidt} and Saskatoon\cite{Sask} and with ChPT
calculations\cite{BKM-photo}. The unpolarized cross sections were
accurately measured and,  despite their small size, the results from
Mainz and Saskatoon are in reasonable agreement.  The experiments were
performed using tagged photons for energies  between threshold (144.7
MeV) and 166 MeV. For the Mainz data there were sufficient statistics
to bin the cross section data in $\simeq$ 1 MeV steps. The ChPT
calculations\cite{BKM-photo} have proven to be quite accurate in
fitting the cross sections with only five empirical low energy
constants at O($p^{4}$)\footnote{Due to correlations in the fitting there are effectively only three independent low energy parameters.}.  In addition, the polarized linear photon asymmetry $\Sigma$ was also measured at Mainz. Here the statistics only allowed us to group the data from threshold to 166 MeV in one cross section averaged energy bin of 159.5 MeV\cite{Schmidt}. The results are shown in Fig.~\ref{fig:photo_thresh}. Here the improvement in the O($p^{4}$) ChPT calculation over the O($p^{3}$) version is seen. This is obtained by fitting the data using the additional low energy constants that appear at  O($p^{4}$). This is an indication of how sensitive this observable is to the small p wave multipoles. Another indication of this is that the dispersion theory calculation, which does agree with the unpolarized cross section data, does not agree with $\Sigma$. This is probably due to a small discrepancy in the $M_{1-}$ multipole which is not well constrained by the other data on which this calculation is based. More recent data taken at Mainz are anticipated to produce five values of $\Sigma$  between threshold and 168 MeV\cite{Hornidge}.

Most of the dynamical models do not accurately predict cross sections for the near threshold $\gamma p \rightarrow \pi^{0} p$ reaction. The exception to this is the DMT model which has accurately predicted the observed cross sections\cite{DMT:threshold,DMT}. However, it does not accurately predict the polarized photon asymmetry $\Sigma$. Again, as a sign of the extreme sensitivity of this observable, when they arbitrarily reduce their $M_{1-}$ amplitude by 15\% they have agreement with the observed value of $\Sigma$ shown in Fig.~\ref{fig:photo_thresh}. However the prediction of this amplitude is not as robust due to the tail of the Roper resonance, vector meson effects, and final state interactions\cite{DMT:threshold,DMT}.

Having discussed the comparison between the calculations and experiment it is of interest to look at the major ingredients of the ChPT\cite{BKM-photo} and DMT\cite{DMT,DMT:threshold} calculations. ChPT employs chiral symmetric Lagrangians with explicit chiral symmetry broken by the quark mass terms. It is an order by order expansion in which unitarity is restored as the order increases. For example, at the tree level unitarity is completely absent, but is mostly restored by the one loop calculations\cite{BKM-photo}. It is gauge invariant and preserves  crossing symmetry. By contrast the 
DMT model  has chiral symmetry in the Lagrangian and is 
unitary to all orders: it uses a pion cloud model for the $\pi$N t matrix which gives good agreement with the $\pi$N phase shifts\cite{Yang:piN}. It enforces gauge invariance but 
   violates crossing symmetry.

A sensitive way to compare theory and experiment is at the level of
the multipoles. Since the observables are bilinear combinations of the
multipoles\cite{obs} this process is often model dependent. However,
in the case of  near threshold photo-pion production an approximate, but
reasonably accurate, model independent multipole extraction is
possible. This is because there are only five real numbers to extract
from the experiments (see e.g. \cite{AB:cusp} for a more detailed discussion). These are the  s wave electric dipole amplitude $E_{0+}$ which is complex, and three p wave amplitudes which are approximately real numbers in this energy region. Due to the low energy theorems of QCD\cite{BKM-photo} (see Eq.~(\ref{eq:E0})) the p wave amplitudes
tend to dominate even relatively close to threshold. The real part of the  s wave
electric dipole amplitude $\Re E_{0+}$ is extracted from the data using
the interference between s and p waves which goes as
$\cos(\theta_{\pi})$ in the differential cross section and leads to
significant errors. The results for  $\Re E_{0+}$  versus  photon energy are
plotted in  Fig.~\ref{fig:photo_thresh}. There is reasonable
agreement between the Mainz and Saskatoon points  as well as with
ChPT\cite{BKM-photo} and the unitary model
calculations\cite{AB:cusp}. The sharp downturn in $\Re E_{0+}$ between the
threshold at 144.7 MeV and the $\pi^{+}$n threshold at 151.4 MeV is
due to a unitary cusp caused by the interference  between the $\gamma
p \rightarrow \pi^{0} p $ and $\gamma p \rightarrow \pi^{+} n $
channels\cite{AB:FW}. The magnitude of the cusp is $\beta= \Re E_{0+}(\gamma p
\rightarrow \pi^{+} n) \cdot a_{cex}(\pi^{+} n \rightarrow \pi^{0} p)$
which is measured to an accuracy of $\simeq$ 30\% from the data
shown\cite{AB:cusp}. The reason for this accuracy limitation is due to
the fact that in addition to the experimental errors in $\Re E_{0+}$, this quantity  is a sum of a (not precisely known) smooth function and a more rapidly varying cusp\cite{AB:FW,AB:cusp}. Therefore it is
important to measure  $\Im E_{0+}$ which starts from close to zero at
the $\pi^{+}n$  threshold energy and rises rapidly as $\beta
p_{\pi^{+}}$. This makes the extraction of $\beta$ as accurate as the
measured asymmetry for $\pi^{0}$ photoproduction from a polarized
target  normal to the reaction plane. We are planning to conduct
future experiments at HI$\gamma$S, a new photon source being
constructed at Duke\cite{HIgS}. These experiments will have full photon
and target polarization and will be a  significant extension of the
results we have at present. 
The estimated error for such an
experiment running at HI$\gamma$S for $\simeq$ 200 hours of
anticipated operation of the accelerator per data point is presented
in  Fig.~\ref{fig:photo_thresh} for $\Re E_{0+}$. There are equally small
error bars estimated for the asymmetry measurement for unpolarized
photons and a transversely polarized proton target. This experiment
will allow us to extract $\Im E_{0+}$. Combining this with an
independent measurement of the $\gamma p \rightarrow \pi^{+} n$ cross
section will allow us to extract $\beta$ at the few \% level and
measure the  charge exchange scattering length $a_{cex}(\pi^{+} n
\rightarrow \pi^{0} p)$ for the first time. We will be able to
compare this to the measured value of $a_{cex}(\pi^{-} p \rightarrow
\pi^{0} n)$\cite{Gotta:atoms} as an isospin conservation test. This
illustrates the power of photo-pion reaction studies with transversely
polarized targets to measure $\pi N$ phase shifts in  completely
neutral  charge channels which are not accessible to pion beam
experiments! This is potentially valuable to help  pin down
experimentally the value of the $\pi N-\sigma$ term which has had a
long, difficult measurement history.

\begin{figure}
\begin{center}
\epsfig{file=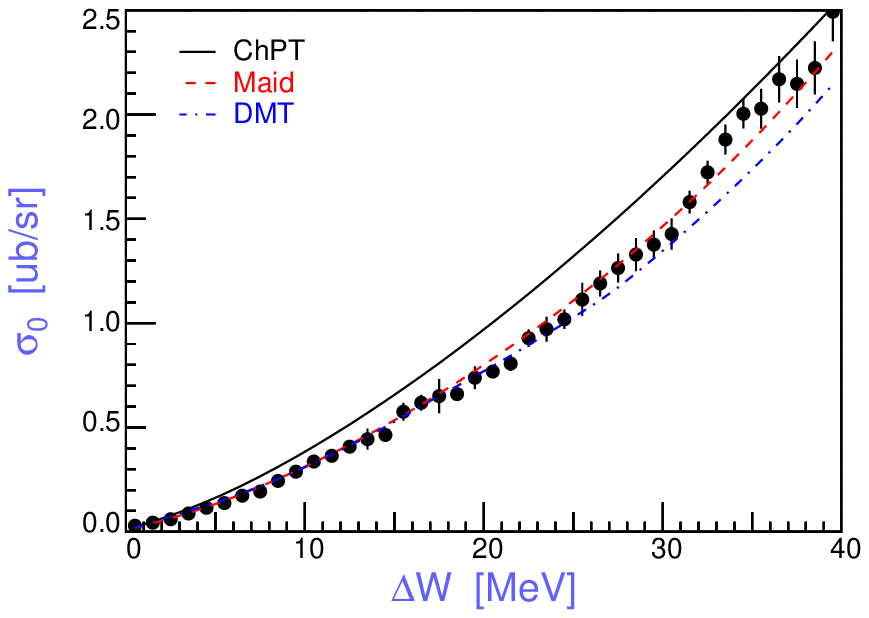, width=2.6in}
\epsfig{file=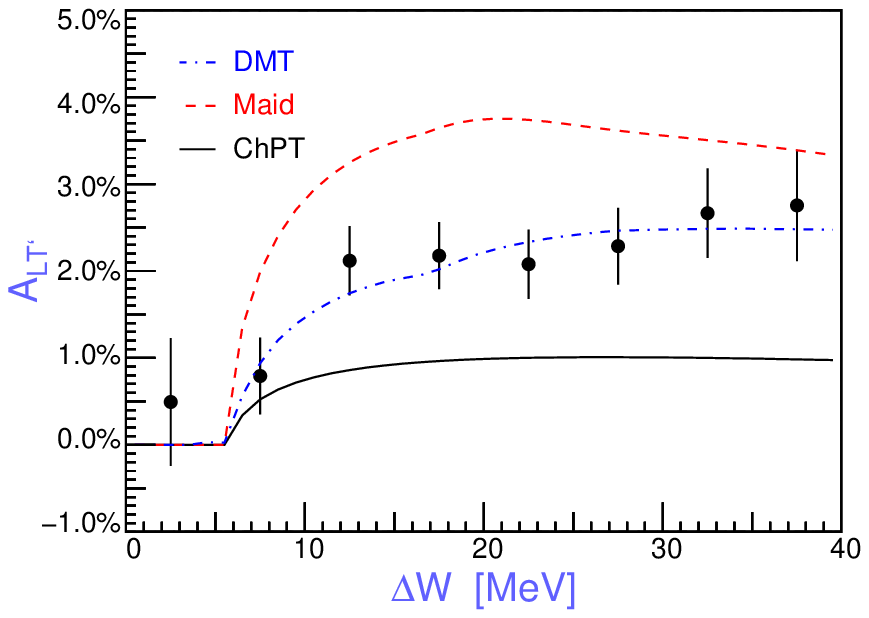,width=2.6in}
\end{center}
\caption{\label{fig:eepi}
Cross section (left panel) and $LT'$ asymmetry (right panel)  for
the $ep \rightarrow e' \pi^{0}p$ reaction at $Q^{2}  = 0.05 {\rm ~GeV}^{2}/{\rm c}^2$ versus $\Delta W$, the center of mass energy above threshold\cite{Weis}. See text for discussion. } 
\end{figure}

Although ChPT has been extremely successful in predicting the cross
sections and the linearly polarized photon asymmetry in the  $\gamma p
\rightarrow \pi^{0} p$ reaction there is a significant discrepancy
with the $e p \rightarrow e' p \pi^{0}$ reaction data  at $Q^{2}
=0.05 {\rm ~GeV}^{2}/{\rm c}^2$ taken at Mainz\cite{Weis}  shown in Fig.~\ref{fig:eepi}. It can be seen that the ChPT calculations\cite{BKM-electro} do not agree with the data although the DMT dynamical model does\cite{DMT:threshold,DMT}. This discrepancy is a potentially serious problem for ChPT which needs to be resolved. The reason is that the present calculations are O$(p^{3}$) and it has been shown that to obtain agreement with the photo-pion data  O($p^{4}$) calculations are needed. 

\begin{figure}
\begin{center}
\epsfig{file=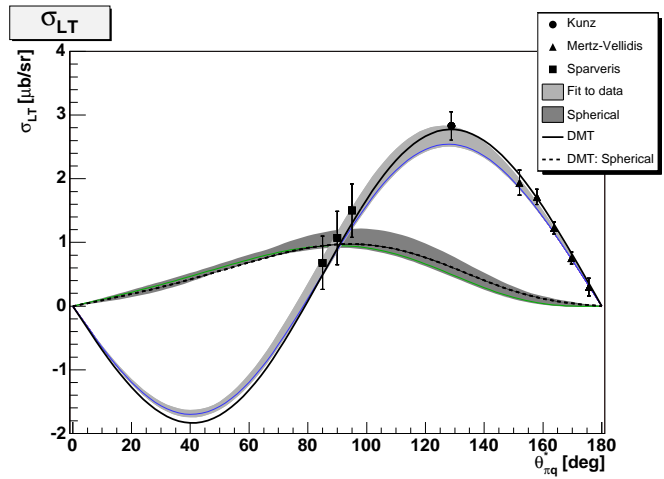, width=3.5in}
\end{center}
\caption{\label{fig:shape}
$\sigma_{LT}$  from the $e p \rightarrow e' \pi^{0} p$ reaction at
  $Q^{2} = 0.127 {\rm ~GeV}^{2}/{\rm c}^2$, $W=1232$ MeV (see
  text). The light grey curve is the fit to the data\cite{nikos} and
  the relatively flat, dark grey curve shows the calculation for the spherical case, i.e. when the quadrupole transition amplitudes are set to zero\cite{SOH,Costas, Stave:fits}. }
\end{figure}

The photo- and electro-pion $\gamma^{*} p \rightarrow \Delta$ reactions
have been extensively used to study non-spherical amplitudes (shape)
in the nucleon and $\Delta$ structure\cite{SOH,AB:overview,PVY}. This is studied by
measuring the electric and Coulomb quadrupole amplitudes  (E2,C2) in
the predominantly magnetic dipole, quark spin flip (M1) $\gamma^{*} N \rightarrow  \Delta$ amplitude. At
low $Q^{2}$ the non-spherical pion cloud is a major contributor to
this (for a review see\cite{SOH,AB:overview,PVY}).
Figure~\ref{fig:shape} shows the Bates data\cite{nikos} for the
transverse-longitudinal  interference cross section
$\sigma_{LT}$ at $Q^{2} =0.127 {\rm ~GeV}^{2}/{\rm c}^2$\cite{nikos}.  This partial cross section is particularly sensitive to the Coulomb quadrupole C2 $\gamma^{*} N \rightarrow  \Delta$ amplitude. This figure shows our best estimate of the difference between
the electro-excitation $\Delta$ for the spherical case (the relatively
flat, dark grey band) and the fit to the data which shows the C2 magnitude\cite{Costas, Stathis,Stave:fits}. The magnitude and $Q^{2}$
evolution of the Coulomb quadrupole amplitude  indicates
that the quark models do not agree with experiment, but that models
with pionic degrees of freedom do\cite{DMT,Sato-Lee}, demonstrating that the crucial
ingredient in the non-spherical amplitude at long range is the pion
cloud. More recently there have been chiral-effective field theory
calculations of this process\cite{PV,GH} which reinforce this observation. As was stated earlier, the presence of long range pionic effects in the non-spherical nucleon and $\Delta$ amplitudes is expected due to the spontaneous hiding of chiral symmetry and the associated p wave pion-nucleon interaction (see Eq.~(\ref{eq:VpiN})). 

\section{Conclusions}

We have shown that the classical isolated $\Delta$ resonance energy dependence (which is well known to be a p wave resonance) also depends on the weakness of the s wave amplitude. This behavior is expected on the basis of the spontaneous hiding of chiral symmetry in QCD: it is observed in the total cross sections for $\pi^{+/-}p$ scattering and in the $\gamma p \rightarrow \pi^{0} p$ reactions. It was shown that in the case of the $\gamma p \rightarrow \pi^{+}n$ reaction, where the Kroll-Ruderman theorem leads to a strong s wave production at threshold, that the energy dependence of the $\Delta$ resonance appears quite different. 

In the well studied, near threshold $\gamma^{*} p \rightarrow
\pi^{0}p$ reaction the agreement between O($p^{4}$) ChPT calculations
and experiment is excellent for the photon data but not so good for
electroproduction at $Q^2=0.05 {\rm ~GeV}^2/{\rm c}^2$ where the calculations have only been carried out to O($p^{3}$), indicating that further work is required. The pion cloud DMT model gives a reasonable description of all of the data.  We have also mentioned further experiments  which will test the theories in a more stringent fashion. These include photo-pion production experiments with transversely polarized targets which have the potential to measure $\pi$N scattering in previously unexplored charge states ($\pi^{0} n$ elastic scattering and charge exchange). If these can be performed with sufficient accuracy they will subject the theory to stringent tests. In addition,  isospin conservation can be checked in a new way.


\begin{thebibliography}{10}

\bibitem{PDB}
Yao, W.-M.  et~al.: J. Phys. {\bf G33}, 1 (2006)

\bibitem{Fermi}
Anderson, H.~L., Fermi, E., Long, E.~A., Nagle, D.~E.:
Phys. Rev. {\bf 85}, 936 (1952)
  
  \bibitem{CD2006}Ahmed, M. W., Gao, H., Holstein,  B., Weller
    H. R.(eds.): Workshop on Chiral Dynamics: Theory and Experiment 2006. To be published by World Scientific, editors, http://www.tunl.duke.edu/events/cd2006/
  
  \bibitem{BM-reviews}
Bernard, V., Meissner, U.-G.:
Submitted to Ann.Rev.Nucl.Part.Sci., [arXiv:hep-ph/0611231 (2006)];
Bernard, V.: [arXiv:hep-ph/0706.0312 (2007)]

\bibitem{AB-CD2006}Bernstein, A. M.: Opening Remarks at Chiral
  Dynamics 2006: Experimental Tests of Chiral Symmetry Breaking,
  [arXiv:hep-ph/0707.4250 (2007)]

\bibitem{DMT}
Kamalov, S. S., Yang, S.~N.: Phys. Rev. Lett. {\bf 83}, 4494 (1999)

\bibitem{DGH}
Donoghue, J.~F., Golowich,  E.~, Holstein, B.~R.: Dynamics of the 
{S}tandard {M}odel.  Cambridge ; New York: Cambridge University Press 1992

\bibitem{Goity:GT}
Goity, J.~L., Lewis, R., Schvellinger, M., Zhang, L.-Z.:
Phys. Lett. {\bf B454}, 115 (1999)

\bibitem{W:pion-scat}
Weinberg, S.: Phys. Rev. Lett. {\bf 17}, 616 (1966)

\bibitem{Gotta:atoms}
Gotta, D.:  Int. J. Mod. Phys. {\bf A20}, 349 (2005)

\bibitem{AB:FW}
Bernstein, A.~M.: Phys. Lett. {\bf B442}, 20 (1998)

\bibitem{BKM-photo}
Bernard, V.~, Kaiser, N., Meissner, U.-G.: Eur. Phys. J. {\bf A11},
209 (2001); Z. Phys. {\bf C70}, 483 (1996); Phys. Lett. {\bf B383}, 116 (1996)

\bibitem{SAID}
Arndt, R. et~al.: Phys. Rev. {\bf C66}, 055213 (2002); http://gwdac.phys.gwu.edu

\bibitem{MAID}
Drechsel, D., Hanstein, O., Kamalov, S.~S., Tiator, L.:
Nucl. Phys. {\bf A645}, 145 (1999); www.kph.uni-mainz.de/MAID

\bibitem{notation}
The notation for the $\gamma^{*} N \rightarrow \pi N$ amplitudes is: the
  capital letter $M,E, L(S)$ stands for the character of the virtual photon,
  magnetic, electric, or longitudinal (or scalar), the subscript stands for
  $l$= orbital angular momentum of the emitted pion, and the $\pm$ indicates
  $j$ = the total angular momentum of the emitted pion = $l$ $\pm$ 1/2.

\bibitem{obs}
Drechsel, D., Tiator, L.:
J. Phys. {\bf G18}, 449 (1992); 
Raskin, A.S., Donnelly, T.W.: Ann.  Phys. {\bf 191}, 78 (1989)

\bibitem{Schmidt}
Schmidt, A. et~al.: Phys. Rev. Lett. {\bf 87}, 232501 (2001)

\bibitem{Sask}
Bergstrom, J.~C. et~al.: Phys. Rev. {\bf C53}, 1052 (1996)

\bibitem{Hornidge}Hornidge, D. : private communication

\bibitem{DMT:threshold}
Kamalov, S.~S., Chen, G.-Y., Yang, S.~N., Drechsel, D., Tiator, L.:
Phys. Lett. {\bf B522}, 27 (2001)

\bibitem{Yang:piN}
Hung, C.-T., Yang, S.~N., Lee, T.-S.~H.: Phys. Rev. {\bf C64}, 034309 (2001)

\bibitem{AB:cusp}
Bernstein, A.~M. et~al.: Phys. Rev. {\bf C55}, 1509 (1997)

\bibitem{HIgS}
http://higs.tunl.duke.edu

\bibitem{dispersion-photo}
Hanstein, O., Drechsel, D., Tiator, L.: Nucl. Phys. {\bf A632}, 561 (1998)

\bibitem{Weis}
Weis, M. et~al.: [arXiv:nucl-ex/0705.3816 (2007)]

\bibitem{BKM-electro}
Bernard, V., Kaiser, N., Meissner, U.-G.:
Nucl. Phys. {\bf A607}, 379 (1996); Phys. Rev. Lett. {\bf 74}, 3752 (1995)

\bibitem{SOH}
Papanicolas, C., Bernstein, A.~M. (eds.): Shape of Hadrons, Athens,
Greece, 27-29 April 2006 AIP Conference
  Proceedings, Volume 904. New York: AIP 2007

\bibitem{AB:overview}
Bernstein, A.~M., Papanicolas,  C.~N.: AIP Conf. Proc. {\bf 904}, 1 (2007)

\bibitem{PVY}
Pascalutsa, V., Vanderhaeghen, M., Yang, S.~N.: Phys. Rept. {\bf 437}, 125 (2007)

\bibitem{nikos}
Sparveris, N.~F. et~al.: Phys. Rev. Lett. {\bf 94}, 022003 (2005)

\bibitem{Costas}
Papanicolas, C.N.: Eur. Phys. J. {\bf A18}, 141 (2003)

\bibitem{Stathis}
Stiliaris, E., Papanicolas,  C.~N.: AIP Conf. Proc. {\bf 904}, 257 (2007)

\bibitem{Stave:fits}
Stave, S., Bernstein, A.~M., Nakagawa, I.: AIP Conf. Proc. {\bf 904}, 245 (2007)

\bibitem{Sato-Lee}
Sato, T., Lee, T.-S. H.: Phys. Rev. {\bf C63}, 055201 (2001)

\bibitem{PV}
Pascalutsa, V., Vanderhaeghen, M.: Phys. Rev. {\bf D73}, 034003 (2006)

\bibitem{GH}
Gail, T.~A., Hemmert, T.~R.: Eur. Phys. J. {\bf A28}, 91 (2006)

\end{thebibliography}

\end{document}